\documentclass[doc,12pt,floatsintext]{apa7}

\usepackage[backend=biber, natbib=true, style=apa]{biblatex}
\DeclareLanguageMapping{american}{american-apa}

\usepackage[T1]{fontenc} 
\usepackage{mathptmx} 
\usepackage{graphicx}%
\usepackage{multirow}%
\usepackage{amsmath,amssymb,amsfonts}%
\usepackage{amsthm}%
\usepackage{mathrsfs}%
\usepackage[title]{appendix}%
\usepackage{xcolor}%
\usepackage{textcomp}%
\usepackage{manyfoot}%
\usepackage{booktabs}%
\usepackage{algorithm}%
\usepackage{algorithmicx}%
\usepackage{algpseudocode}%
\usepackage{listings}%
\usepackage{bm}%
\usepackage{longtable}%
\usepackage{hyperref}
\usepackage{makecell}
\usepackage{forest}
\usepackage{tikz}

\title{Capturing the Complexity of Human Strategic Decision-Making with Machine Learning} 
\shorttitle{Capturing Game Complexity} 
\author{Jian-Qiao Zhu$^{1}$, Joshua C. Peterson$^{2}$, Benjamin Enke$^{3}$, Thomas L. Griffiths$^{1,4}$}

\affiliation{\normalsize{$^{1}$Department of Computer Science, Princeton University}\\
\normalsize{$^{2}$Faculty of Computing and Data Science, Boston University}\\
\normalsize{$^{3}$Department of Economics, Harvard University, and NBER}\\
\normalsize{$^{4}$Department of Psychology, Princeton University}}

\abstract{Understanding how people behave in strategic settings--where they make decisions based on their expectations about the behavior of others--is a long-standing problem in the behavioral sciences. We conduct the largest study to date of strategic decision-making in the context of initial play in two-player matrix games, analyzing over 90,000 human decisions across more than 2,400 procedurally generated games that span a much wider space than previous datasets. We show that a deep neural network trained on these data predicts people's choices better than leading theories of strategic behavior, indicating that there is systematic variation that is not explained by those theories. We then modify the network to produce a new, interpretable behavioral model, revealing what the original network learned about people: their ability to optimally respond and their capacity to reason about others are dependent on the complexity of individual games. This context-dependence is critical in explaining deviations from the rational Nash equilibrium, response times, and uncertainty in strategic decisions. More broadly, our results demonstrate how machine learning can be applied beyond prediction to further help generate novel explanations of complex human behavior.}


\begin{document}
\maketitle 

\begin{figure}[t!]
    \centering
    \includegraphics[width=\textwidth]{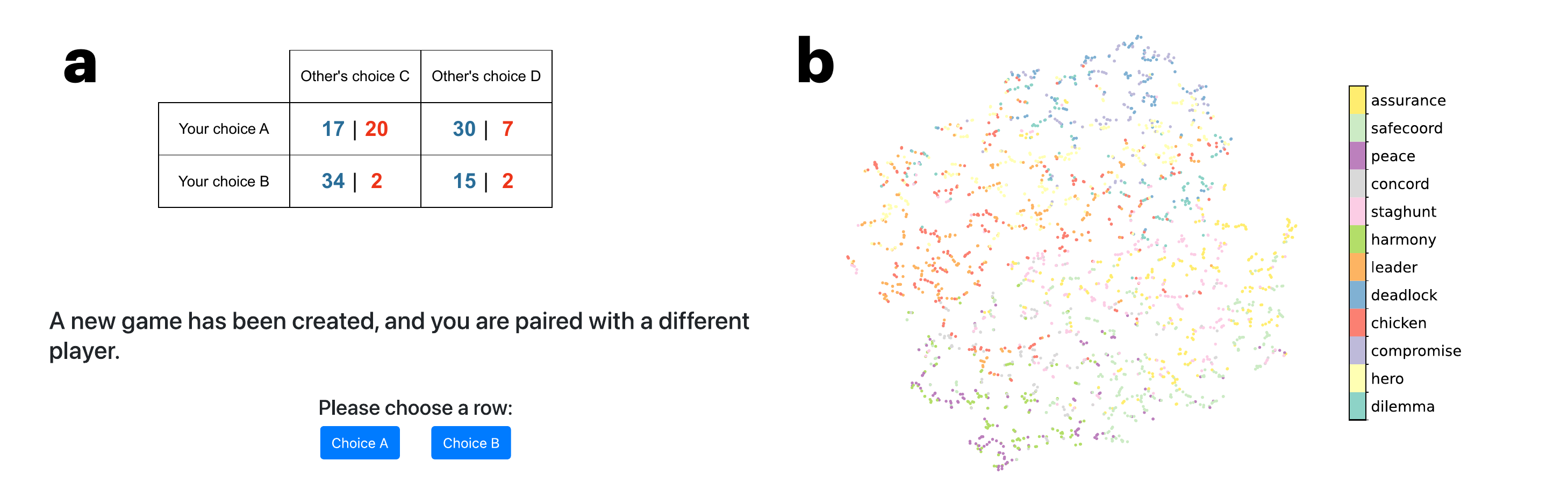}
    \caption{ \textbf{Matrix games.} 
             \textbf{(a)} An example game interface presented to participants, who acted as the row player in each $2 \times 2$ game. The blue numbers represent the payoffs for the row player, who chooses between strategies $A$ and $B$, while the red numbers represent the payoffs for the column player, who chooses between strategies $C$ and $D$.
             \textbf{(b)} Visualization of game space. Each game is uniquely represented by an 8-integer vector, corresponding to the payoffs to the two players under different configurations of choices. We used $t$-distributed stochastic neighbor embeddings \citep{van2008visualizing} to visualize the spatial relationship between games in a 2D plot, using the Euclidean distance between the embeddings of our best-performing neural network model. Points represent individual games. The colors represent the game topology specific to the row player following \textcite{robinson2005topology}.
             }
    \label{fig:intro_to_games_and_models}
\end{figure}

Strategic decision-making is essential when people's outcomes depend on both their own and other people's actions. As a consequence, it is an important topic in various disciplines within the social sciences, including economics, psychology, political science, and artificial intelligence, as well as cultural and biological evolution \citep{vonneuman1944, camerer2011behavioral, rawls1971atheory, baker2017rational, fehr1999theory, wright2017predicting}. The most widely studied type of game in the social sciences is the class of $2 \times 2$ matrix games (see Figure~\ref{fig:intro_to_games_and_models}a), which have been found to illuminate behavior in contexts including, among many others, human cooperation and the evolution of morality \citep{bloom2012religion}, price setting and production decisions by firms \citep{fudenberg1991game}, the coordination of investment decisions \citep{bell1988game}, and the positioning of political candidates \citep{morrow1994game}.

The rational model of strategic decisions -- the Nash equilibrium -- is based on two key assumptions: mutual consistency in beliefs about opponents’ strategies and mutual rationality in best responding to those beliefs. However, despite the prevalent use of Nash equilibria in analyzing matrix games, research has shown that human players often violate both of these assumptions \citep{mckelvey1995quantal, camerer2011behavioral, gachter2004behavioral, weizsacker2003ignoring}. Consequently, the effectiveness of these equilibria in explaining people's strategic behaviors is limited \citep{mckelvey1992experimental, mckelvey1995quantal}. This has prompted the development of behavioral game theory, which has identified  various extensions and refinements that produce a closer match to human decisions \citep{eyster2019errors, fehr1999theory, fudenberg2019predicting, golman2020dual}.

Despite a proliferation of behavioral models, evaluating the performance of these models has relied on relatively small datasets based on a select group of games, even when combined across different datasets and papers \citep{wright2017predicting, fudenberg2019predicting}. As a result, it remains unclear how well the most popular models of strategic decision-making perform in general. For instance, even seemingly ``simple'' types of strategic interaction can differ widely in the cognitive difficulty they pose for the actors, yet our understanding of how game complexity shapes behavior is limited. To explore these questions, we conducted a large-scale study of strategic behavior by densely sampling the enormous space of $2\times 2$ game structures. We use the resulting dataset to assess the explanatory power of leading behavioral models, quantifying their prediction performance against a machine learning model trained on the same data. This strategy allows us to identify systematic variation that is not captured by existing models, leading us to develop a new interpretable model that captures human behavior almost as well as the machine learning algorithm.

We procedurally generated a dataset of 2,416 matrix games involving monetary gains (see Figure~\ref{fig:intro_to_games_and_models}b), significantly expanding the diversity of game scenarios studied in prior datasets (a 17-fold increase in the number of games tested relative to the largest meta-analysis in the literature \citep{wright2017predicting}). 
To systematically sample game matrices, our game generation algorithm is based on the Robinson and Goforth topology for $2\times 2$ games, which is constructed using ordinal order graphs of payoffs \citep{robinson2005topology}. Each player has 12 unique order graphs in their respective payoff matrices, representing the different ways to rank payoffs between their two strategies. Consequently, there are $12\times 12=144$ possible game types, considering each player's order graph independently. We populated all types with at least one pure-strategy Nash equilibrium  with procedurally-generated game matrices (see Supplementary Information for details). To investigate human behavior in these settings, we recruited 4,900 participants via Prolific, each of whom was instructed to participate in 20 distinct games, sampled randomly without replacement from the pool of procedurally-generated games. In total, 93,460 strategic decisions were recorded. No feedback was provided to participants between games, and the players were randomly rematched after each game. Thus, the observed behaviors can be interpreted as initial game play strategies.

\begin{figure}[t!]
    \centering
    \includegraphics[width=\textwidth]{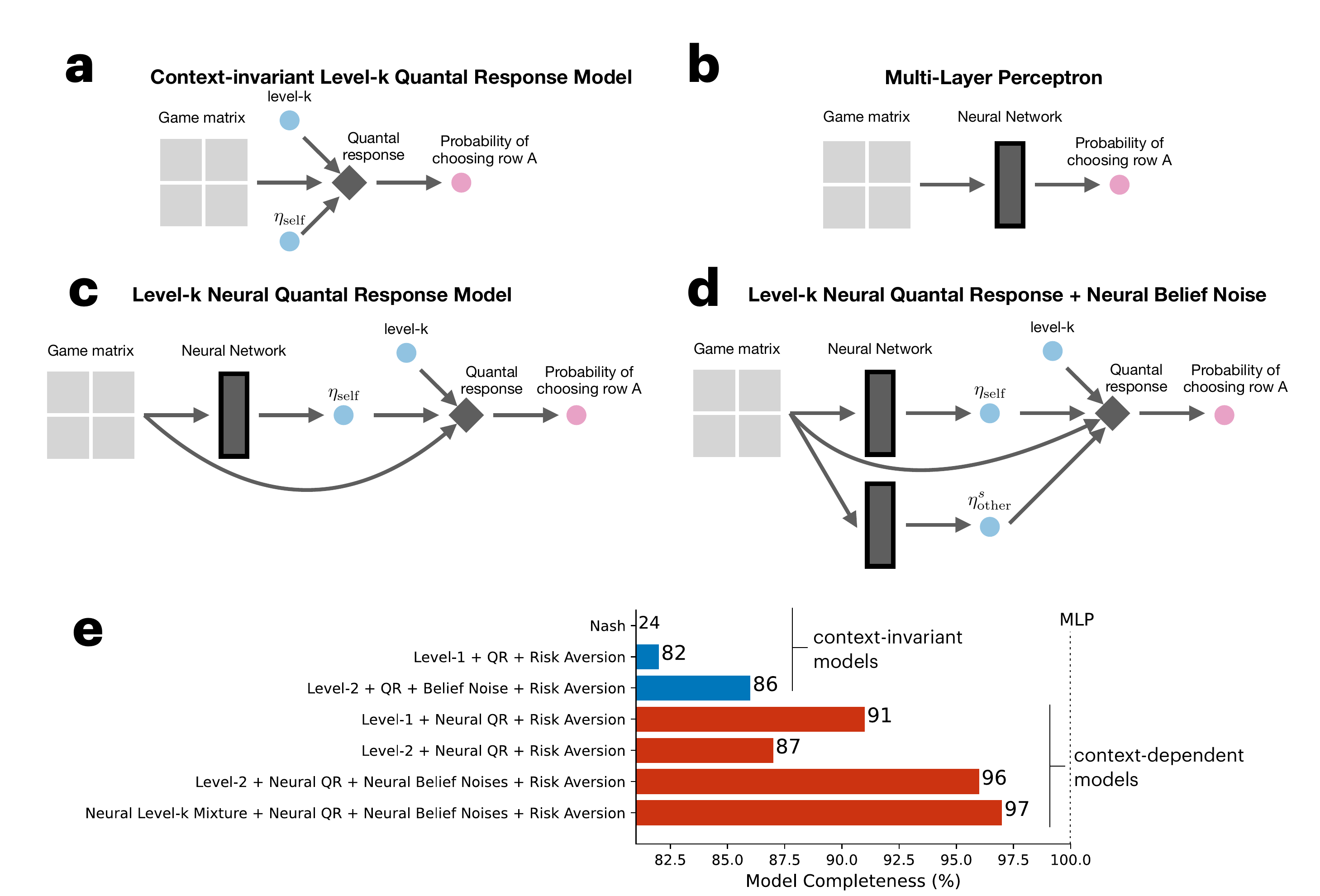}
    \caption{ \textbf{Model comparisons.} 
             \textbf{(a)} The context-invariant level-$k$ quantal-response model involves three parameters that do not vary across games: strategic sophistication (i.e., $k$), the players' noisiness (i.e., $\eta_\text{self}$) and risk aversion. 
             \textbf{(b)} A Multi-layer Perceptron (MLP) model directly uses the game matrix as input to estimate choice probabilities, without imposing any specific game-theoretic decision-making structure.
             \textbf{(c)} The level-$k$ neural quantal-response model is a context-dependent model allowing the $\eta_\text{self}$ parameter to vary across games. It employs an MLP model, which uses the game matrix as input to estimate game-specific $\eta_\text{self}$.
             \textbf{(d)} The level-$k$ neural quantal-response and neural belief noise model extends the model in (c) by further learning the game-specific $\eta^s_\text{other}$ and $k$ parameters through two MLP models, each of which takes the game matrix as input. 
            \textbf{(e)} Context-dependent models, incorporating at least one neural network component that allows some or all model parameters to vary across games, outperform context-invariant models in terms of completeness. Higher completeness indicates greater predictive accuracy for human behaviors, with 100\% completeness matching the predictive accuracy of the MLP model. All reported results were based on 10-fold cross-validation (see Supplementary Information for details). Our focus was on the heterogeneity across games rather than the heterogeneity across participants. 
             }
    \label{fig:model_completeness}
\end{figure}

We used the resulting dataset to evaluate various models of strategic decision-making with a train-validation split. We fit all models by minimizing the mean squared error (MSE) between model predictions and empirical choice frequencies at the game level on the training set, and then assess their performance in a validation set (see Supplementary Information for details).

The models are based on three key insights from the behavioral game theory literature, each representing a well-established aspect of human cognition in strategic interactions. First, the players may have limited strategic sophistication and compute the best-response function only a small number of times, as captured in level-$k$ models and similar approaches \citep{camerer2004cognitive,crawford2013structural}. Second, a player may exhibit noise in their decision making and also take into account the noisiness of others' behavior, as in quantal-response equilibrium models \citep{mckelvey1995quantal}. Third, players may be risk averse, meaning they prefer less uncertainty \citep{murnighan1988risk,fudenberg2019predicting}.

All models that we estimate are variants of our baseline behavioral model, which is a risk-averse level-$k$ quantal response (QR) model (see Figure~\ref{fig:model_completeness}a) \citep{wright2017predicting}. According to this model, a player forms beliefs about their opponent's behavior (albeit possibly imperfectly) and selects the risk-averse best response based on these beliefs (albeit possibly imperfectly). 
Two interrelated equations characterize game play under these assumptions. The first equation describes the expected utility from any given strategy, and the second describes how expected utility translates into behavior. When the row player in a matrix game has strategies $A$ and $B$ available and the column player decides between $C$ and $D$ (see Figure \ref{fig:intro_to_games_and_models}a), the quantal-response function asserts that the row player's probability of choosing $A$ is an increasing function of the difference in expected utility (EU) between $A$ and $B$, where the inverse of $\eta_\text{self}$ governs the player's noisiness:
\begin{align}
    p(A) & = \frac{1}{1+e^{-\eta_\text{self} [EU(A) - EU(B)]}}
\end{align}
The expected utility of any given strategy, in turn, is given by computing the utility of the payoffs ($x$), weighted by the row player's subjective belief, $p^s$, about whether the column player plays $C$ or $D$:
\begin{align}
    EU(A) &= p^s(C|k,\eta^s_\text{other}) \; U(x_{A,C}) \; + \; p^s(D|k,\eta^s_\text{other}) \; U(x_{A,D}) \;,
\end{align}
where $x_{i,j}$ represents the payoff for the row player when they choose row $i$ and the column player chooses column $j$. In these equations, the key insights from the behavioral game theory literature appear in three different components. First, in assessing the expected utility from any given strategy, limited strategic sophistication (captured by $k$) affects a player's expectation of their opponent's play and, hence, expected utility. Second, the expectation of the opponent's play also depends on subjective beliefs about the noisiness of the other player ($\eta^s_\text{other}$). Following the literature, we initially assume that $\eta_\text{self}=\eta^s_\text{other}\equiv \eta$, meaning that each player believes that their opponent is as noisy as they are \citep{mckelvey1995quantal}. Third, expected utility takes into account risk aversion, parameterized by constant absolute risk aversion, $U(x)=(1-e^{-\alpha x})/\alpha$.

We quantify the performance of each model by benchmarking it against two extremes. First, to set a lower performance bound, we employed a random model, which predicts choice probabilities uniformly at random. Second, to set an upper performance bound, we employ a deep neural network (Multi-Layer Perceptron, MLP) that uses the game matrix as input and targets empirical choice frequencies as its output. The performance of all other models was assessed based on their \textit{completeness}, which measures how well a model approximates the neural network upper bound from a starting point of random play \citep{fudenberg2022measuring, wright2017predicting}. For example, a completeness of 50\% indicates that a model’s predictive accuracy is exactly in between those of the random model and the MLP. To determine overall model completeness, we averaged the completeness values calculated from two predictive accuracy metrics, MSE and $R^2$.

We found that allowing for limited strategic sophistication (level-$k$), quantal-response noise and risk aversion has large effects on model fit (see Figure~\ref{fig:model_completeness}e). The standard Nash equilibrium achieves only 24\% completeness. A model that combines level-1 thinking, quantal-response noise and risk aversion achieves 82\% completeness. While these results illustrate that the success of behavioral game theory also translates into the much larger space of games that we analyzed, they also reveal substantial room for improvement of the structural decision-making model relative to the deep neural network. To close this gap, we considered what might be missing from existing theoretical accounts of strategic behavior.

A fundamental characteristic of existing models in the behavioral game theory literature is their context-invariance: they are uniformly applied and estimated, with identical parameters, irrespective of the characteristics of the game. However, evidence from related decision-making domains, such as choice between risky lotteries \citep{peterson2021using, enke2023quantifying}, suggests that the parameters of behavioral models can be highly context-sensitive, for example because the complexity or cognitive difficulty of decisions varies across problems. The deep neural network model represents an extreme case of potential context-dependence because it is  capable of forming its predictions based on the specific characteristics of each game. Yet while this model achieves high prediction accuracy, it is less useful for understanding human strategic decision-making because it is an uninterpretable ``black box,'' lacking the instructive format of structural decision-making models.

To build a bridge between these two approaches, we progressively introduced context-dependence into the structural decision-making models, by systematically substituting the structural parameters of behavioral models with a neural network that is responsive to game-specific features. We focus on three key behavioral primitives: (i) the level of strategic sophistication (i.e., $k$), (ii) the player’s level of noisiness (i.e., $\eta_\text{self}$; neural quantal response model), and (iii) the player’s beliefs regarding the noisiness of others (i.e., $\eta^s_\text{other}$; neural belief noise model). We allowed each of these structural parameters to be endogenously determined by the game matrices through a neural network, thus introducing context-dependence into the structural decision-making models in a disciplined and interpretable manner. To capture the context-dependence of $k$ using neural networks, we let the neural network model the distribution of $k$: $p(k)$. As a result, the strategic choice predicted by this model can be interpreted as a mixture of players operating at different levels of $k$ (See Supplementary Information for details).

As an example, consider the level-$k$ neural quantal response model (Figure \ref{fig:model_completeness}c), in which the player's noisiness was predicted by a neural network. In this approach, an MLP was used to model the function $\eta_\text{self} = f_{MLP}(\text{game matrix})$, allowing the $\eta_\text{self}$ parameter to vary across games. Once trained, the MLP can predict $\eta_\text{self}$ for any given game matrix. By combining the strategic sophistication level $k$ with the MLP-predicted $\eta_\text{self}$, the level-$k$ quantal-response function iteratively applies the quantal-response function $k$ times to produce the probability of the row player choosing row $A$ for a given game matrix. Similarly, for all other neurally-augmented behavioral models, the neural networks effectively learn a function that maps the game matrix to the parameters in the original behavioral model. Overall, the procedure of augmenting behavioral models with neural networks intuitively captures the possibility that players' level of strategic sophistication or their noisiness is not fixed but, instead, depends on features of the game.

As illustrated in Figure~\ref{fig:model_completeness}e, integrating neural networks (i.e., MLPs) into the level-$k$ quantal-response model to account for context-dependence significantly enhances model completeness. When neural networks replaced all three structural parameters, model completeness reached 97\%. The second-best model, achieving a completeness of 96\%, maintained strategic sophistication at a fixed level of $k=2$, yet allows for context-dependence in $\eta_\text{self}$ and $\eta_\text{other}^s$. This context-dependence (i.e., game-specific noisiness) suggests an important role for decision-making difficulty -- in some games, it is considerably easier to identify one's best response than in others, even fixing a given level of strategic sophistication. Moreover, by comparing models that vary solely in $\eta_\text{self}$ or $\eta_\text{other}^s$ across different games (see Table \ref{tab:mse_r2_completeness}), we found that variations or context dependence in $\eta_\text{self}$ play a more significant role. This finding suggests that context dependence has a greater impact on individuals’ ability to optimize their own responses than on their ability to infer the likely actions of others.

\begin{figure}[t!]
    \centering
    \includegraphics[width=\textwidth]{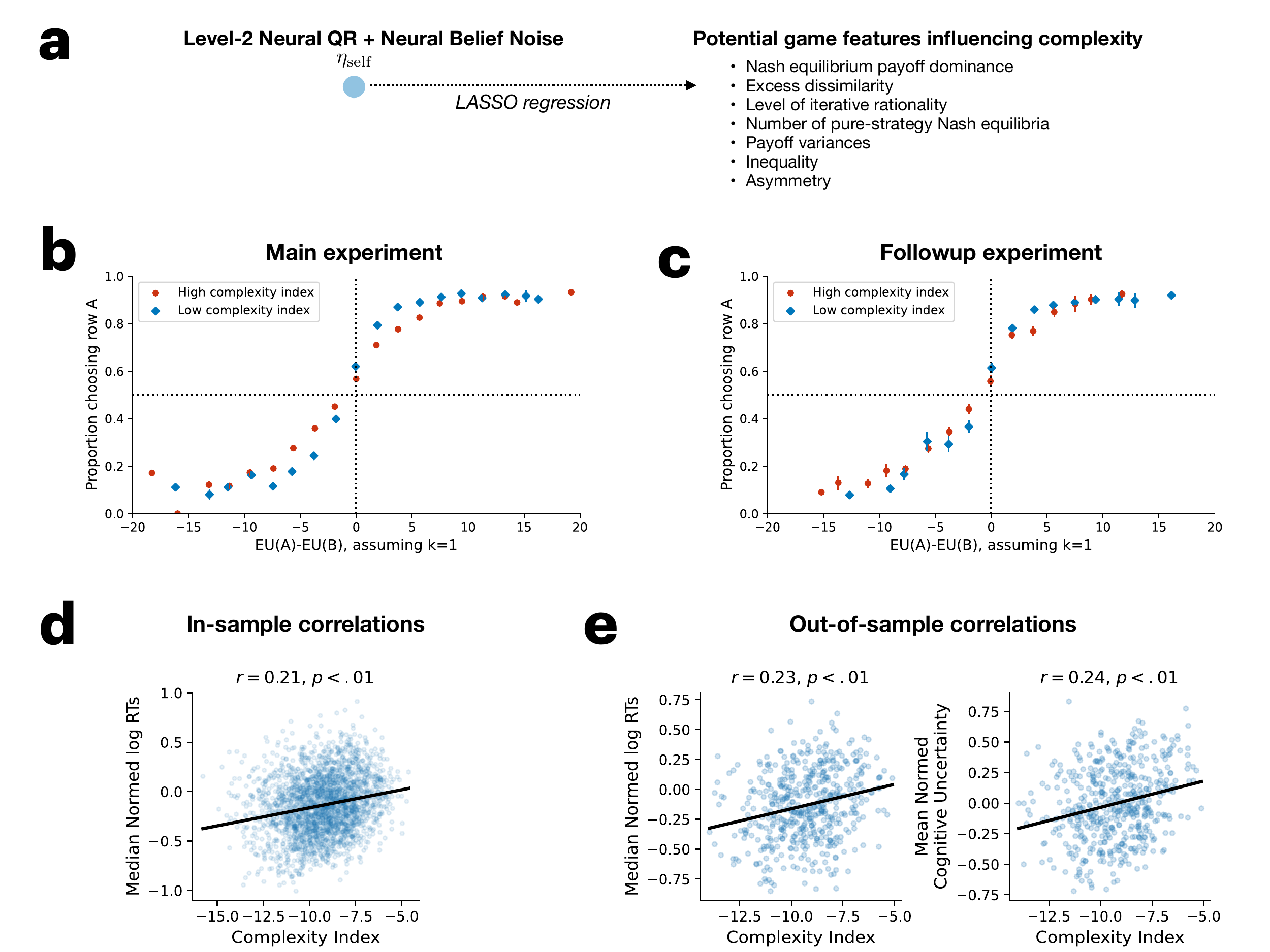}
    \caption{\textbf{Developing an interpretable complexity index for strategic games.}
    \textbf{(a)} To construct an index of game complexity, we use LASSO regressions to learn  game features that correlate with the game-specific $\eta_\text{self}$ parameter that is estimated by the MLP in the Level-2 Neural QR and Neural Belief Noise model. 
    \textbf{(b)} The psychometric functions illustrate the relationship between the expected utility differences of two strategies and the proportion of choices for strategy $A$. Expected utility was calculated under the assumption of a level-1 player. Red (blue) dots represent high (low) complexity games, determined by a median split on the complexity index. Error bars represent $\pm$SE.
    \textbf{(c)} The same psychometric function and effect of complexity was found in the followup experiment.
    \textbf{(d)} The complexity index shows a statistically significant correlation with response times (RTs) in the games of the main experiment. 
    \textbf{(e)} The complexity index generalized to the followup experiment and demonstrated statistically significant correlations with both RTs and cognitive uncertainty ratings. 
    }
    \label{fig:complexity_index}
\end{figure}

To build intuition for why context-dependent noisiness has large effects on model completeness, Figure~\ref{fig:complexity_index}b shows the game-level link between the share of subjects choosing strategy $A$ and the difference in expected utility between strategies $A$ and $B$, as estimated from the level-2 neural quantal response model. We split the sample by the median estimated noisiness ($\eta_\text{self}$). We see a strong attenuation pattern: for games with below-median noisiness, the link between strategy choice frequencies and expected utility estimates is considerably more compressed. Our level-2 neural quantal response model, a context-dependent model, achieves higher completeness both because it learns in which types of games players are more likely to best-respond, and because it learns in which types of games players are more likely to identify the best response of their opponent (given the presumed level of strategic sophistication, $k$).


The high performance of our context-dependent models indicates that both the ability of players to optimally respond to their beliefs and their capacity to reason about others' behaviors are contingent on the specific game being played. We attribute this sensitivity to \textit{game complexity}, by which we mean the cognitive difficulty of (i) forming beliefs about other players' strategies and (ii) optimally responding to these beliefs. To quantitatively define and validate a metric of game complexity, we first developed a composite index of game complexity, and then validated it by showing correlations with independent markers of complexity in a preregistered second experiment.

We leveraged our unusually large and diverse set of games to extract some of the specific game features that drive complexity, and aggregated them into a sparse and interpretable index of game complexity. To this effect, we analyzed the predicted $\eta_\text{self}$ values from the context-dependent level-2 quantal response model (see Figure~\ref{fig:complexity_index}a: the Level-2 Neural QR + Neural Belief Noise model) and conducted a LASSO regression on a large set of structural game features (see Supplementary Information for detailed definitions). This analysis identified a concise set of influential game features. Some features are prominent in the literature, such as the number of steps of iterative reasoning required for equilibrium choices. Other features are more novel, including (i) a measure of the cognitive difficulty of navigating tradeoffs across different strategies, akin to work on dissimilarity and tradeoff complexity in the literature on lottery choice \citep{enke2023quantifying,shubatt2024similarity}; (ii) the variance and scale of payouts \citep{peterson2021using}; and (iii) the inequality and asymmetry in payouts between players.

We collapsed these interpretable game features into a composite index of complexity (see Supplementary Information). This index can be structurally interpreted as capturing the magnitude of the negated $\eta_\text{self}$ predicted by game features. Because this index is defined based on objective game features, it can be readily computed by other researchers in any standard dataset.

A first piece of evidence that our complexity index indeed captures the difficulty of strategic decision-making is that, in our main experiment, the index has shown a positive correlation with response times (Pearson's $r=0.21, p<.01$, see Figure~\ref{fig:complexity_index}d), indicating that individuals tend to spend more time thinking in games that we classify as more complex.

To reinforce and broaden these within-sample correlations and our interpretation of the index, we conducted a preregistered follow-up experiment testing a new set of 500 games on a new cohort of participants. This experiment adhered closely to the procedure of the main experiment, with the sole modification being that, after each game, participants were required to report their cognitive uncertainty (in percentage terms) about whether the strategy they selected is actually their best decision \citep{enke2023cognitive}. The results confirmed that the index robustly predicts out-of-sample behavioral outcomes. As shown in Figure~\ref{fig:complexity_index}e, we replicated the positive correlation between RTs and the complexity index (Pearson's $r=0.23, p<.01$). Additionally, we observed a positive correlation between cognitive uncertainty and the complexity index (Pearson's $r=0.24, p<.01$), suggesting that participants tend to exhibit higher cognitive uncertainty in their strategic choices when faced with more complex games. Finally, in this follow-up experiment, we also studied the ability of our complexity index to predict behavioral attenuation in strategic choice. Figure \ref{fig:complexity_index}c shows that the link between strategy choice frequencies and estimated expected utility differences is again considerably more compressed in the more complex problems. These findings confirm that our complexity index can generalize to out-of-sample strategic decisions.


Large-scale experiments and machine learning techniques have significantly aided our exploration of the vast space of cognitive mechanisms underlying strategic choices. Our findings reveal that a player’s own noisiness in responding to an opponent’s behavior, as well as their beliefs about the opponent’s noisiness, are critical factors in determining initial game play. Moreover, the degrees of noise vary across different games, displaying significant context dependence. To further understand the game features that contribute to this context dependence, we developed a complexity index that quantifies a player’s noisiness. This index is interpretable and readily generalizable to other matrix games, as it is based solely on features derived from the game matrix. The follow-up experiment confirmed that the complexity index effectively captures various aspects of human behavior in matrix games with differing levels of complexity. These results illustrate the promise of large-scale experiments and machine learning methods in furthering our understanding of strategic decision-making, in particular given the emerging body of theoretical work on complexity in behavioral game theory.

\section*{Acknowledgments}
This work and related results were made possible with the support of the NOMIS Foundation. We thank Nick Chater and Shengwu Li for helpful discussions.

\printbibliography

\pagebreak

\begin{appendices}

\setcounter{figure}{0}
\renewcommand{\thefigure}{S\arabic{figure}}
\setcounter{table}{0}
\renewcommand{\thetable}{S\arabic{table}}
\setcounter{equation}{0}
\renewcommand{\theequation}{S\arabic{equation}}

\begin{center}
    {\large Supplementary Information}
\end{center}

\section*{Materials and Methods}

\subsection*{Main Experiment}

\subsubsection*{Game generation algorithm}

The \texttt{games2p2k} dataset comprises 2,416 instances of 2$\times$2 normal-form games and a total of 93,460 strategic choices made by 4,673 participants. Each game is uniquely identified by its payoff matrix, an 8-element vector encapsulating all payoff details for both row and column players. All payoffs are represented as integers and are restricted to a range of 1 to 50. Consequently, the set of all possible 2$\times$2 games encompasses $50^8$ games before considering permutations of game matrices. 

To generate games from this space, we employed a random generation process to create 8-item payoff matrices, where each item represents a random draw from a two-tiered uniform distribution $U[1, u]$ where $u\sim U[1,50]$. Next, we excluded games that lacked pure-strategy Nash equilibria and categorized each remaining game using Robinson and Goforth’s topology for 2$\times$2 games \citep{robinson2005topology}. This process results in a diverse set of 142 distinct game types. Given the prevalence of dominance games, which constituted 87.5\% of the game types, we selectively reduced their representation in our dataset. Specifically, we generated 3 instances for each double-dominance game, 8 instances for each single-dominance game, and 22 instances for each non-dominance game, culminating in a comprehensive collection of 1,208 games. Given that each game is played by both the row and column players, our dataset comprises a total of 2,416 (i.e., 1,208$\times$2) game instances.

\subsubsection*{Participants}
We recruited a total of 4,900 participants via the Prolific Academic platform, out of which 4,673 individuals (1,942 males, 1,782 females, and 949 who opted not to disclose their gender) successfully completed the 10-minute experiment. These participants ranged in age from 18 to 86, with a median age of 37. Participants were required to be from the US and have completed at least 100 submissions with a 95\% acceptance rate or higher on the Prolific Academic platform. Participants were guaranteed a base monetary compensation of \$2.00 (hourly rate of \$12.00), with the possibility of an additional bonus up to \$0.50 contingent on the result of a randomly selected game (1 point equals to \$0.01). The experimental sessions were carried out in Dec 2023. Informed consent was obtained from all participants (Princeton University IRB number 10859: ``Computational Cognitive Science'', and Harvard University IRB number 16-1753).

\subsubsection*{Procedure}
The experiment was programmed using Dallinger 9.11.0. Before the beginning of the main experiment, participants were provided with instructions and an example of a normal-form game matrix. To ensure comprehensive understanding of the game representation, participants were required to correctly answer a pre-experiment multiple-choice quiz. Following this, participants were presented with a sequence of 20 one-shot games in normal form. All game presentations were adapted to a row player perspective, thereby making all participants choose between Row \textit{A} and Row \textit{B}, while the specifics of the game and the role they assumed were recorded in the background. Participants only played a game once. As the primary objective of our task was to observe choices in one-shot games, no practice or coaching period was provided.

To mitigate learning effects and preclude reputation building, each trial involved a different game and a new opponent for every participant. That is, participants were anonymously and randomly paired for each game. Feedback pertaining to a participant's performance in a specific game was withheld. The only occasion where feedback was provided occurred post-completion of the entire experimental session, indicating the bonus earned by the participant. The sequence in which the games were presented was randomly varied across participants. Moreover, for every game, the rows and the columns of the payoff matrix have equal chances being swapped, resulting in a total of 4 possible permutations for a game. The bonus was determined by randomly selecting one game played by the participant and their corresponding opponent in that game.

\subsection*{Follow-up Experiment}

This experiment was preregistered at \url{https://osf.io/xrvaw/}.

\subsubsection*{Games}
We employed the identical game generation algorithm used in the main experiment to create a set of 500 new normal-form games, maintaining similar proportions of dominance games as observed in the main experiment.

\subsubsection*{Participants}
We recruited another 1,013 participants from the Prolific Academic platform, of which 1,008 (346 males, 416 females, and 246 who opted not to disclose their gender) successfully completed the 10-minute experiment. Participant ages ranged from 18 to 88 years, with a median age of 35. The same filter used in the main experiment were applied here: participants were required to be from the U.S. and to have completed at least 100 submissions with a 95\% acceptance rate or higher. As in the main experiment, participants received a base compensation of \$2.00 (hourly rate of \$12.00), with the possibility of earning a bonus of up to \$0.50 from a randomly selected game. The experimental sessions were conducted in April 2024.  Informed consent was obtained from all participants (Princeton University IRB number 10859: “Computational Cognitive Science”, and Harvard University IRB number 16-1753).

\subsubsection*{Procedure}
The follow-up experiment replicated the exact procedure of the original study, with the sole modification that after participants made a strategic choice for a game, they were presented with a secondary question: ``How certain are you that choosing [$X$] is actually your best decision?'' where $X$ denotes the selected option. Note that this query remained concealed until after a choice had been made. Participants were then required to express their confidence in their decision by adjusting a slider ranging from 0\% (least confident) to 100\% (most confident). The slider's thumb was also hidden until the participant interacted with it by clicking on the slider bar.

\section*{Data Preprocessing}

We collected three types of behavioral measures from the two experiments (see Table~\ref{tab:behave_measures} for an overview): strategic choices, response times (RTs) in making these choices, and subjective certainty ratings (confidence) associated with the choices. We aggregated data at the trial level to derive game-level behaviors. The empirical choice frequency, calculated across different participants for the same game, was used as the probability of choosing a particular strategy in that game. This measure served as the target variable for subsequent model fittings. RTs were initially log-transformed: $\hat{RT} = \ln RT$. Subsequently, we normalized these log-transformed RTs within participants by subtracting each participant's mean $\hat{RT}$ and dividing by the standard deviation of their $\hat{RT}$. The median of these normalized $\hat{RT}$ values was then used to represent the game-level RTs. Confidence judgments were normalized within each participant, and the mean of these normalized confidence values was calculated to represent game-level confidence.

\begin{table}[h!]
    \centering
    \caption{Overview of Behavioral Measures Collected in the Main and Follow-up Experiments.}
    \begin{tabular}{lcc} \hline
        Behavioral measures & Statistical properties & Recording experiments \\ \hline
        Choices & Binary outcomes & Both experiments \\
        Response Times & Non-negative integer values & Both experiments \\
        \makecell[l]{Confidences\\(Negated Cognitive Uncertainty)} & Continuous range $[0,1]$ & Follow-up experiment \\ \hline
    \end{tabular}
    \label{tab:behave_measures}
\end{table}

\section*{Model Details}

\subsection*{Context-invariant Models}
The first class of models for human strategic choices comprises context-invariant models. These are parametric models with free parameters that remain constant across different games, hence the term ``context invariance.'' This consistency in parameters, however, does not imply that the models will predict identical behaviors for the same set of parameters across various games.

\textbf{Nash Equilibrium in Pure and Mixed Strategies.}
In matrix games, a pure-strategy Nash equilibrium (PSNE) is an outcome where each player chooses a single action (or strategy) and no player can benefit by changing their action, given that the other players’ actions remain the same. In other words, every player’s choice is their best response to the choices of the other players. By contrast, a mixed-strategy Nash equilibrium (MSNE) occurs when players randomize over possible actions according to certain probabilities, and no player can improve their expected payoff by unilaterally changing their own probability distribution over actions. In MSNE, each player’s mixed strategy is the best response to the mixed strategies of the other players.


\textbf{Level-$k$ Model.}
The level-$k$ model in behavioral game theory explains strategic choices by assuming varying degrees of strategic sophistication among players \citep{mauersberger2018levels, costa2001cognition, nagel1995unraveling}. Level-0 players select their actions without strategic consideration, often randomly or according to a fixed rule, and they do not attempt to predict the actions of others. Level-1 players, assuming that all other players are at level 0, make decisions that are best responses to the expected actions of level-0 players. Each player at level $k$ selects their action based on the belief that others are at level $k-1$. Thus, the level-$k$ model captures a hierarchy of strategic reasoning, where each successive level represents a deeper anticipation of others’ behavior based on the preceding level’s actions.

\textbf{Quantal Response Equilibrium.}
In a Nash equilibrium, each player’s strategy is the best response to the strategies of the other players, and no player can improve their outcome by unilaterally changing their strategy. Conversely, the quantal response model offers a more realistic perspective by incorporating the idea that players may not always make perfectly rational decisions \citep{rosenthal1989bounded, mckelvey1995quantal}. Instead of always choosing the best response, players select strategies with probabilities that increase with the expected payoff, introducing a noisy response. This means that players are more likely to choose better strategies, but they can still sometimes make suboptimal choices.

In the quantal response equilibrium (QRE), players are not assumed to be perfectly rational. Rather, they respond to expected payoffs in a probabilistic manner. The probability of selecting a particular strategy increases with the expected payoff of that strategy, yet suboptimal choices remain possible. Specifically, we used a logit quantal response function \citep{mckelvey1995quantal}. Each player’s strategy is influenced by the probability distribution of the other players’ strategies, leading to a more flexible and realistic equilibrium concept.

\textbf{Level-$k$ Quantal Response Model.}
While the level-$k$ model employs best response functions and the QRE model utilizes equilibrium solutions, the level-$k$ quantal response (QR) model integrates both approaches \citep{stahl1994experimental, golman2020dual}. In this model, level-$k$ players assume that all other players are at level $k-1$, and they make decisions that are \textit{quantal}, rather than best, responses to the expected actions of level-$k-1$ players. Furthermore, it is important to note that level-$k$ players expect level-$k-1$ players to apply the same quantal response logic to level-$k-2$ players, continuing this pattern until reaching the random actions of level-$0$ players. In essence, level-$k$ players believe that level-$k-1$ players respond noisily to the actions of players one level below them, mirroring their own quantal response function.

\textbf{Level-$k$ Quantal Response and Belief Noise Model.}
We also considered a simple extension of the Level-$k$ QR model by assuming that players might not believe their opponents exhibit the same level of noisiness as themselves, namely \textit{Belief Noise}. There is empirical evidence supporting this assumption where, for example, people tend to underestimate the rationality of their opponents in strategic choices \citep{weizsacker2003ignoring}. Specifically, players maintain two sets of quantal response functions simultaneously: one for themselves when they need to respond quantally to their opponents’ choices, and another set reflecting what they believe their opponents use when responding to their own choices.



\subsection*{Context-dependent Models}

The other class of models exhibits context dependence, meaning that the parameter values of these models change in response to different game matrices. This context dependence is implemented using a Multilayer Perceptron (MLP) that processes game matrices as inputs and generates corresponding model parameters.

\textbf{Multilayer Perceptron.}
The neural networks employed in this study use a consistent MLP architecture, featuring 3 hidden layers with 300 neurons each. Sigmoid functions are applied for all nonlinear activations, except when the network output is a probability, in which case a softmax function is used. The highest performance in modeling the \texttt{games2p2k} dataset was achieved by an MLP that directly predicts choice probabilities from the game matrix input:
\begin{align}
    p(A) = f_{MLP}(\text{game matrix})
\end{align}

\textbf{Neural Network Augmented Behavioral Models.}
We focus on adapting three key behavioral primitives using MLPs. First, we make the level of strategic sophistication (i.e., $k$) dependent on the game matrices. However, since levels are discrete and non-differentiable variables, we extend this concept by assuming a distribution $p(k)$ of participants at each level $k$, where $k=0,1,2,3$. This approach effectively creates a mixture model that represents a range of Level-$k$ players. Next, the player’s own level of noisiness in response to opponents’ choices (i.e., $\eta_\text{self}$) and the player’s beliefs about the noisiness of their opponents (i.e., $\eta_\text{other}$) are modeled using two additional MLPs. The outputs of these MLPs are then input into a quantal response function to generate choice probabilities.


\subsection*{Optimization and Cross-validation}
We optimized all models by minimizing the mean squared errors (MSE) between the predicted outcomes and the actual proportions of $A$ (or ``up'') choices made by row players in each game. For models incorporating neural networks, they were implemented using the Python libraries \texttt{jax} \citep{jax2018github} and \texttt{haiku} \citep{haiku2020github}. The \texttt{games2p2k} dataset was partitioned randomly into training (80\%), validation (10\%), and testing (10\%).

\textbf{Neural network training details.}
Models involving neural networks were trained in batches of 64 games each. The dataset was also augmented through permutations. For each game, three additional duplicates were created by switching the rows, switching the columns, and switching both the rows and columns. The choice probabilities were then adjusted accordingly. Training was terminated when validation errors increased consecutively over two evaluation intervals of 100 epochs. The performance of the trained models was assessed on the testing set, with a particular emphasis on evaluating their generalization capabilities. The learning rate was fixed at $10^{-3}$, and the Adam optimizer was employed \citep{kingma2014adam}.

\textbf{Context-invariant models training details.}
Models without neural network components underwent optimization using the Nelder-Mead method via the \texttt{minimize} function of the \texttt{Scipy} library \citep{2020SciPy-NMeth}. Training these models involves randomly splitting the dataset into training (90\%) and testing (10\%) sets. The optimization phase concluded upon the convergence of the Nelder-Mead algorithm on the training set. Model performance was also evaluated on the testing set.

The cross-validation for both context-invariant and context-dependent models was repeated a total of 10 rounds for all models. Different random partition over games was performed in each round. Both the mean and standard error for the MSE and the $R^2$ were recorded for the testing set (see Table \ref{tab:mse_r2_completeness}).

\begin{table}[h!]
    \centering
    \caption{Comparisons Between Context-invariant and Neurally-Augmented Context-dependent Models}\vspace{1mm}
    \begin{tabular}{llll}\hline
        Model & MSE (SE) $\downarrow$ & $R^2$ (SE) $\uparrow$ & Completeness (\%) $\uparrow$ \\ \hline
        Random & .0875  & .0003  & 0 (lower bound) \\ 
        Nash & .1625  & .2234  & 24 \\
        L1+QR & .0255 (.0006) & .7134 (.009) & 77.5 \\
        L1+QR+Risk & .0218 (.0005) & .7509 (.0092) & 82 \\
        L2+QR & .026 (.0006) & .6968 (.0077) & 76.5\\
        L2+QR+Risk & .0244 (.0005) & .7306 (.0056) & 79\\
        L2+QR+Belief+Risk & .0182 (.0005) & .7949 (.0062) & 86\\
        L3+QR & .0275 (.0006) & .6868 (.0072) & 75\\
        L3+QR+Risk & .0238 (.0005) & .7303 (.006) & 79\\
        L3+QR+Belief+Risk & .0181 (.0006) & .7906 (.0063) & 86.5\\
        QRE & .0269 (.0007) & .7112 (.0099) & 76.5\\
        QRE+Risk & .0251 (.0005) & .7213 (.0061) & 78\\
        QRE+Belief+Risk & .0183 (.0005) & .7888 (.006) & 86\\
        \hline
        L1+\textcolor{orange}{QR} & .0229 (.0011) & .773 (.0091) & 82.5 \\
        L1+\textcolor{orange}{QR}+Risk & .0145 (.0006) & .8361 (.0077) & 91 \\
        L2+\textcolor{orange}{QR} & .0206 (.0012) & .78 (.0095) & 84\\
        L2+\textcolor{orange}{QR}+Risk & .0178 (.0006) & .7968 (.0065) & 87\\
        L2+\textcolor{orange}{QR}+Belief+Risk & .0125 (.0003) & .862 (.0045) & 94 \\
        L2+QR+\textcolor{orange}{Belief}+Risk & .0291 (.0006) & .6895 (.0073) & 74 \\
        L2+\textcolor{orange}{QR}+\textcolor{orange}{Belief}+Risk &  .0109 (.0003) & .8784 (.0045) & 96\\
        L3+\textcolor{orange}{QR} & .0389 (.0089) & .6944 (.0259) & 68.5 \\
        L3+\textcolor{orange}{QR}+Risk & .0187 (.0003) & .7856 (.0036) & 85.5 \\
        L3+\textcolor{orange}{QR}+Belief+Risk & .0124 (.0002) & .8615 (.0024) & 94 \\
        L3+QR+\textcolor{orange}{Belief}+Risk & .032 (.001) & .6776 (.0091) & 71.5 \\
        L3+\textcolor{orange}{QR}+\textcolor{orange}{Belief}+Risk &  .0119 (.0003) & .8674 (.0045) & 94\\
        \textcolor{orange}{QR}E & .0245 (.0011) & .7492 (.0096) & 80 \\
        \textcolor{orange}{QR}E+Risk & .0188 (.0005) & .7904 (.0067) & 86 \\
        \textcolor{orange}{QR}E+Belief+Risk & .0125 (.0005) & .8604 (.0058) & 94 \\
        QRE+\textcolor{orange}{Belief}+Risk & .0282 (.0009) & .7231 (.0081) & 76.5 \\
        \textcolor{orange}{QR}E+\textcolor{orange}{Belief}+Risk &  .0116 (.0004) & .8732 (.0036) & 95\\
        \textcolor{orange}{L}+QR+Risk & .0194 (.0003) & .7834 (.0052) & 85\\
        \textcolor{orange}{L}+\textcolor{orange}{QR}+\textcolor{orange}{Belief}+Risk & .0096 (.0003) & .8931 (.0039) & 97 \\
        MLP & .0073 (.0001) & .9194 (.0022) & 100 (upper bound)\\
        \hline
    \end{tabular} \\ \vspace{1mm}
    \textit{Note.} The model completeness was determined by averaging the completeness values calculated from Mean Squared Error (MSE) and $R^2$, except for the Nash equilibrium model, where completeness was solely based on $R^2$. Colorized texts denote components of the model that are implemented as neural networks. Specifically, \textcolor{orange}{QR} indicates that $\eta_\text{self}$ is a neural network, \textcolor{orange}{Belief} signifies that $\eta_\text{other}^s$ is a neural network, and \textcolor{orange}{L} means that a neural network predicts a mixture of level-$k$ players. The numbers in parentheses represent the standard errors from the 10-fold cross-validation.
    \label{tab:mse_r2_completeness}
\end{table}

\section*{Developing a Game Complexity Index}

In our computational modeling, we identified the player's level of noisiness (i.e., the $\eta_\text{self}$ parameter from context-dependent models) as an effective target for deriving an interpretable index that reflects perceptions of game complexity. This choice is based on the premise that game complexity is likely linked to deviations from rational behavior, which are captured by the $\eta_\text{self}$ parameter. To develop this index, we employed a two-step procedure: First, we reviewed existing game features known in the literature and introduced a series of novel features. This process created a comprehensive set of candidate game features that potentially elucidate aspects of game complexity. Subsequently, we refined this set to identify the most informative features by employing LASSO regression, using the $\eta_\text{self}$ parameter as the dependent variable and the full suite of game features as independent variables.

\begin{figure}[h!]
    \centering
    \includegraphics[width=0.5\textwidth]{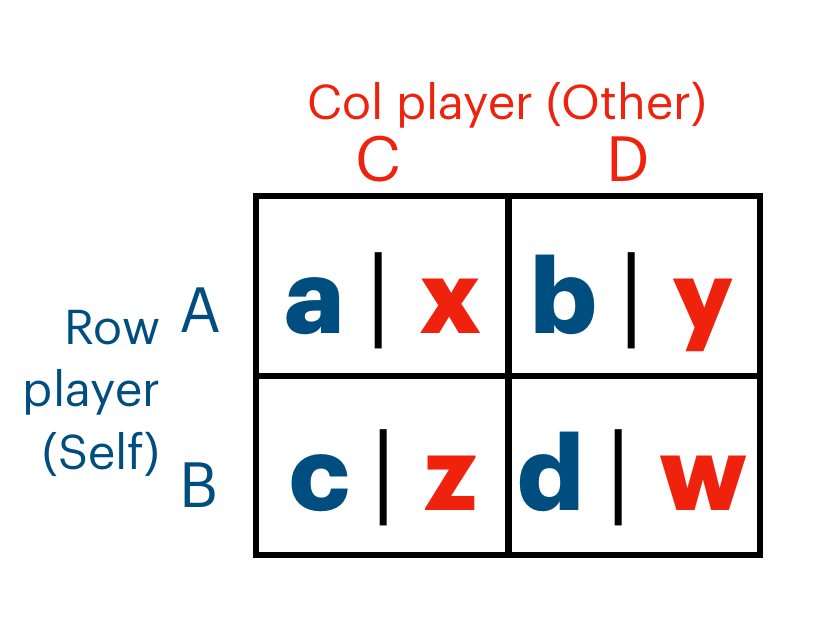}
    \caption{Example of a game matrix: The payoff matrix for the row player (`self') is in blue, while the payoff matrix for the column player (`other') is in red.}
    \label{fig:SI_example_game}
\end{figure}

\subsection*{Game Topology}
Using the illustrative game depicted in Figure \ref{fig:SI_example_game}, we summarize Robinson and Goforth’s topology for $2\times 2$ matrix games \citep{robinson2005topology}. Since both row and column players have 12 unique order graphs in the payoff space of their respective payoff matrices, this topology is constructed on a set of $12\times 12=144$ ordinal order graphs. For instance, from the perspective of a row player, the 12 order graphs including Chicken (or Hawk-Dove, Snowdrift, $c>a>b>d$), Leader (or Battle of the sexes, $c>b>a>d$), Hero ($c>b>d>a$), Compromise ($c>d>b>a$), Deadlock (or Altruist's Dilemma, $c>d>a>b$), Prisoner's Dilemma ($c>a>d>b$), Stag Hunt (or Trust, $a>c>d>b$), Assurance ($a>d>c>b$), Safe Coordination ($a>d>b>c$), Peace (or Club, $a>b>d>c$), Harmony ($a>b>c>d$), Concord ($a>c>b>d$). The same topology was applied to the column player’s payoff matrix.

\subsection*{Game Features}

In this section, we provide a detailed description, along with examples, of the interpretable features of a game matrix. We also discuss their expected correlation with game complexity.

\paragraph{Dominant Solvability.}

Dominant solvable games, which means that a game is solvable by iterated deletion of dominated action, are generally considered easier to play because the payouts from one strategy consistently outperform all others. Players' abilities to identify the optimal strategy only require the capacity to best respond. This feature can be independently defined for both the player (row player) and the opponent (column player). For instance, considering the illustrative game depicted in Figure \ref{fig:SI_example_game}, we define the dominant solvability for both players (equality can occur in only one of these inequalities):
\begin{align}
    \text{DominantSolvable}_\text{self} & = 
    \begin{cases} 
        1 & \text{if } a\geq c \text{ and } b\geq d \text{ and not (} a=c \text{ and } b=d \text{)}\\
        1 & \text{if } a\leq c \text{ and } b\leq d \text{ and not (} a=c \text{ and } b=d \text{)}\\
        0 & \text{otherwise}
    \end{cases}  \\
    \text{DominantSolvable}_\text{other} & = 
    \begin{cases} 
        1 & \text{if } x\geq y \text{ and } z\geq w \text{ and not (} x=y \text{ and } z=w \text{)}\\
        1 & \text{if } x\leq y \text{ and } z\leq w \text{ and not (} x=y \text{ and } z=w \text{)}\\
        0 & \text{otherwise}
    \end{cases}  
\end{align}

\paragraph{Excess Dissimilarity.}

Excess dissimilarity, initially introduced in \citep{enke2023quantifying} to understand complexity in risky choices, quantifies the difficulty of aggregating values across states. In risky choices, the measure calculates payout differences between states for two lotteries, then deducts the difference in expected values. We expand this measure to two-player matrix games, further assuming that the opponent's strategies are random. Using the game illustrated in Figure \ref{fig:SI_example_game}, we compute the excess dissimilarity indices for both players as follows:
\begin{align}
    \text{Dissimilarity}_\text{self} & = \frac{|a-c|}{2}+\frac{|b-d|}{2}-|\mu_\text{up}-\mu_\text{down}| \\
    \text{Dissimilarity}_\text{other} & = \frac{|x-y|}{2}+\frac{|z-w|}{2}-|\mu_\text{left}-\mu_\text{right}|
\end{align}
where $\mu_\text{up}=\frac{a+b}{2}, \mu_\text{down}=\frac{c+d}{2}, \mu_\text{left}=\frac{x+z}{2}, \mu_\text{right}=\frac{y+w}{2}$.

\paragraph{Levels of Iterative Rationality.}

We define the levels of iterative rationality as the minimum level $k$ at which the best-response dynamic converges to a fixed strategy. For $2\times 2$ games, we set the upper limit of $k$ to 3 due to the possibility of infinite, cyclical best-response dynamics.

\paragraph{Number of Nash Equilibria.}

We analyze pure-strategy and mixed-strategy Nash equilibria separately (PSNE and MSNE) and determine the number of equilibrium strategies for a game matrix.

\paragraph{Nash Equilibrium Payoff Dominance.}
We compute whether each pure-strategy Nash equilibrium (PSNE) in a game yields the maximum possible payouts for both players. If at least one such equilibrium exists, we classify the game as exhibiting Nash equilibrium payoff dominance; otherwise, this feature is assigned a value of zero. Moreover, we compute Non-Nash Equilibrium Payoff Dominance, which is defined as the presence of an outcome that payoff-dominates all other outcomes but is not a PSNE.

\paragraph{Nash Equilibrium Pareto Dominance.} 
Following \citep{fudenberg2019predicting}, we also assess whether a PSNE offers superior payouts compared to all other PSNEs. If such a PSNE exists, we classify the game as demonstrating Nash equilibrium Pareto dominance. If the game possesses only a single PSNE, it is automatically categorized as Pareto dominant. All other cases are classified as false.

\paragraph{Pure Motives.}
The notion of pure motives was examined in \citep{devetag2008playing}. In games governed by pure motives, players' choice preferences are expected to exhibit perfect rank correlation. This includes scenarios with positive correlations, such as coordination games, as well as negative correlations, like zero-sum games. Games featuring both antagonistic and coordination motives, such as the chicken game or the prisoner's dilemma, are said to be mixed motives games.

\paragraph{Max Payouts.}
For a given game matrix, we calculate the maximum potential payouts for each player. Referring to the example game in Figure \ref{fig:SI_example_game}, we define the maximum payouts as follows:
\begin{align}
    \text{Max}_\text{self} & = \max \{ a, b, c, d\} \\
    \text{Max}_\text{other} & = \max \{ x, y, z, w\} 
\end{align}

\paragraph{Payoff Variances.}
Variations in payouts may influence strategy selection. Here, we employ the canonical variance measure. To illustrate, let's examine the game matrix in Figure \ref{fig:SI_example_game}:
\begin{align}
    \text{PayoffVar}_\text{self} &= \frac{1}{4}\Big[(a-\mu_\text{up})^2+(b-\mu_\text{up})^2 +(c-\mu_\text{down})^2+(d-\mu_\text{down})^2\Big] \\
    \text{PayoffVar}_\text{other} &= \frac{1}{4}\Big[(x-\mu_\text{left})^2+(z-\mu_\text{left})^2 +(y-\mu_\text{right})^2+(w-\mu_\text{right})^2\Big]
\end{align}

\paragraph{Deviations from Zero-Sum Games.}

Zero-sum games exhibit a distinct payout structure where gains for one player directly correspond to losses for another. To gauge deviations from a standard zero-sum game, we introduce a measure. First, we compute the absolute sum of payout differences for each player when jointly deviating from an outcome. Subsequently, we aggregate these absolute sums across all four possible deviations within a $2 \times 2$ game. Illustrated through the example game in Figure \ref{fig:SI_example_game}, we calculate this metric as outlined below:
\begin{align}
    \text{NonZeroSum} = |a-c+x-z| + |a-b+x-y| + |c-d+z-w| + |b-d+y-w|
\end{align}

\paragraph{Inequality in Payouts.}
Social preferences may also play a role in shaping strategic decisions \citep{fehr1999theory}. To capture specific aspects of these preferences, the following features have been devised. Inequality in payouts assesses the disparity in the maximum potential payouts players could attain from the game. In the context of the game depicted in Figure \ref{fig:SI_example_game}, this can be computed as follows:
\begin{align}
    \text{Inequality} = \max\{a,b,c,d\} - \max\{x,y,z,w\}
\end{align}

\paragraph{Asymmetry in Payouts.}
Another source of payout inequality arises from variations in payouts when both players employing a similar strategy. This can be quantified as the extent of deviations from a symmetric game, where payouts would be identical if both players adopted the same strategy. To compute the asymmetry in payouts for the game depicted in Figure \ref{fig:SI_example_game}, we proceed as follows:
\begin{align}
    \text{Asymmetry} = \frac{1}{4}\Big[ |a-x| + |b-z| + |c-y| + |d-w| \Big]
\end{align}

\subsection*{Expected Relationships Between Game Features and Complexity}

We have now developed a set of game features that characterize a game matrix and are relevant to game complexity. Here, we present their predicted correlations with complexity and provide justifications for these expectations (see Table \ref{tab:game_feature_definition}). Broadly, these justifications can be summarized into the following categories.

First, games with strategies that players will never choose, as long as they are best responding to \textit{any} belief, are considered less complex. Conversely, other types of games require rational choices that ensure both consistency in beliefs about opponents’ strategies and the ability to best respond to those beliefs. Dominant Solvability is the feature with this property.

Second, game features that facilitate coordination are expected to reduce game complexity. Efficiency, defined by payoff dominance or Pareto dominance \citep{schelling1980strategy,harsanyi1995new}, is considered separately from rational equilibrium solutions. When equilibrium solutions coincide with efficient solutions, players can coordinate on the overlapping outcome, making the game easier to solve. Conversely, when different solution concepts do not align, the game becomes more challenging. Game features with this property include Nash Equilibrium Payoff and Pareto Dominance, Maximum Payouts, and Non-Nash Equilibrium Payoff Dominance.

Third, a game is generally expected to be less complex if players can easily adopt their opponent’s perspective \citep{schelling1980strategy}. This occurs when players’ preferences are perfectly rank-correlated, either positively (e.g., coordination games) or negatively (e.g., zero-sum games). The following game features capture this property: Pure Motives, Deviation from Zero-Sum Games, Inequality in Payouts, and Asymmetry in Payouts.

Fourth, games that require greater cognitive resources to process are intuitively more complex. For instance, choosing among multiple equilibria increases complexity, as reflected by the number of Pure Strategy Nash Equilibria (PSNE) and Mixed Strategy Nash Equilibria (MSNE). Moreover, the computation of expected values can vary in complexity, leading to the inclusion of features such as Excess Dissimilarity and Payoff Variance. Finally, iterative thinking steps needed to form consistent beliefs and choose an equilibrium strategy should also matter for players with bounded cognitive resources. This concept is captured by the Levels of Iterative Rationality.

\begin{table}[h!]
    \centering
    \caption{Developed game features, both existing and novel, and their expected correlation with game complexity.}
    \begin{tabular}{lcc} \hline
       Game feature  & \makecell{Expected correlation with\\game complexity} \\ \hline
        DominantSolvable$_\text{self}$   & - \\
        DominantSolvable$_\text{other}$  & - \\
        Nash Equilibrium Payoff Dominance & - \\
        Nash Equilibrium Pareto Dominance & - \\
        Pure Motives & - \\
        Max$_\text{self}$   & - \\
        Max$_\text{other}$   & - \\
        Deviation from Zero-Sum Games  & - \\
        \hline
        Levels of Iterative Rationality  & + \\
        Dissimilarity$_\text{self}$  & + \\
        Dissimilarity$_\text{other}$  & + \\
        Number of PSNE   & + \\
        Number of MSNE   & + \\
        Non-Nash Equilibrium Payoff Dominance & + \\
        PayoffVar$_\text{self}$  & + \\
        PayoffVar$_\text{other}$  & + \\
        Inequality in Payouts  & + \\
        Asymmetry in Payouts  & + \\
       \hline
    \end{tabular}\\ \vspace{1mm}
    \textit{Note.} Negative correlations suggest that the presence or higher values of certain game features correspond to a decrease in game complexity, while positive correlations indicate the opposite.
    \label{tab:game_feature_definition}
\end{table}

\subsection*{Selection of Game Features}

\textbf{Decision tree regressions.}
To identify important features, we applied decision tree regression using interpretable game features directly on the choice probabilities in \texttt{games2p2k}. By fixing the maximal tree depth to 3, we found that the most important feature was whether the game had a Nash Equilibrium that payoff-dominates all other outcomes. This was followed by excess dissimilarity of the player's own payoff matrix and levels of iterative rationality, which appeared in the second layer of the decision tree. The third layer was omitted as it contained repetitions of the features in the second layer, suggesting that these three features are critical in deciding strategic choices.

\vspace{2mm}

\begin{center}
\begin{forest}
  for tree={
    draw,
    rounded corners,
    align=center,
    top color=white,
    bottom color=white,
    edge+={->},
    font=\small,
  }
  [Nash Equilibrium Payoff Dominance.
    [Dissimilarity$_\text{self}$ $\leq2.5$, edge label={node[midway, left]{False}}
    ]
    [Levels of iterative rationality $\leq1$, edge label={node[midway, left]{True}}
    ]
  ]
\end{forest}
\end{center}

\vspace{1mm}

\textbf{LASSO regressions.}
Considering the array of game features that may impact strategic decisions, we develop a game complexity index to explain the noisiness in players' strategic choices. This index is designed to encapsulate deviations from rationality observed within our games. First, we normalized the game features across different games to ensure comparability among game features. Subsequently, we implemented a LASSO regression by setting the multiplier of the $L_1$ term at $0.2$, targeting the negated $\eta_\text{self}$ parameter from our best-fitting level-2 quantal response model, with the normalized game features serving as independent variables. Results were summarized in Table \ref{tab:lasso_results}.

\begin{table}[h!]
    \centering
    \caption{The results of LASSO regressions.}
    \begin{tabular}{lll} \hline
        Game feature & $-\eta_\text{self}$ & $-\eta_\text{other}^s$ \\ \hline
        $\text{DominantSolvable}_\text{self}$ & . & . \\
        $\text{DominantSolvable}_\text{other}$ & . & . \\
        $\text{Dissimilarity}_\text{self}$ & 0.28 & .\\
        $\text{Dissimilarity}_\text{other}$ & . & 0.08\\
        LevelIterRational & 0.38  & .\\
        NumPSNE & . & .\\
        NumMSNE & .  & . \\
        PayoffDomEquilibrium & -0.80 & . \\
        PayoffDomNonEquilibrium & .  & . \\
        ParetoDomEquilirium & . & .\\
        PureMotives & . & .\\
        $\text{Max}_\text{self}$ & 0.30 & . \\
        $\text{Max}_\text{other}$ & . & 0.01\\
        $\text{PayoffVar}_\text{self}$ & 0.40 & .\\
        $\text{PayoffVar}_\text{other}$ & . & .\\
        NonZeroSum & .  & .\\
        Inequality & 0.85 & .\\
        Asymmetry & -0.09  & .\\
         \hline
         Intercept & -9.28 & -1.09\\
         $R^2$ & 0.33 & 0.02\\
         \hline
    \end{tabular}  \\
    \label{tab:lasso_results}
\end{table}

\section*{Robustness Check}

To further validate our modeling results, we trained both context-invariant and context-dependent models using data from the main experiment and tested their performance using data from the follow-up experiment. For the context-invariant models, training was conducted on the entire main experiment dataset. For the context-dependent models, we applied an optional stopping rule to terminate neural network training by partitioning the main experiment data into training and validation sets with a 90/10 split. The remaining training details were consistent with those previously described. As shown in Table \ref{tab:robustness_check}, our main results were successfully replicated.

\begin{table}[h!]
    \centering
    \caption{Comparisons of models trained on the main experiment and tested on the follow-up experiment.}\vspace{1mm}
    \begin{tabular}{lll}\hline
        Model & MSE (SE) $\downarrow$ & $R^2$ (SE) $\uparrow$ \\ \hline
        Nash & .1680  & .2561 \\
        L1+QR+Risk & .0203  & .758  \\
        L2+QR+Belief+Risk & .016  & .8113  \\
        L3+QR+Belief+Risk & .0167  & .8026  \\
        QRE+Belief+Risk & .0166  & .8039  \\
        \hline
        L1+\textcolor{orange}{QR}+Risk & .0128 (.0001) & .8547 (.0032) \\
        L2+\textcolor{orange}{QR}+Belief+Risk & .0109 (.0002) & .875 (.0027) \\
        L2+\textcolor{orange}{QR}+\textcolor{orange}{Belief}+Risk & .0102 (.0001) & .8868 (.0018) \\
        L3+\textcolor{orange}{QR}+Belief+Risk & .0108 (.0002) & .8798 (.0022) \\
        L3+\textcolor{orange}{QR}+\textcolor{orange}{Belief}+Risk & .0104 (.0001) & .8855 (.0018) \\
        \textcolor{orange}{QR}E+Belief+Risk & .0111 (.0002) & .877 (.0029) \\
        \textcolor{orange}{QR}E+\textcolor{orange}{Belief}+Risk & .0105 (.0001) & .8835 (.0018) \\
        \textcolor{orange}{L}+\textcolor{orange}{QR}+\textcolor{orange}{Belief}+Risk & .0081 (.0002) & .9065 (.0023) \\
        \hline
        MLP & .0073 (.0001) & .9194 (.0022) \\
        \hline
    \end{tabular}\\ \vspace{1mm}
    \textit{Note.} Colorized texts denote components of the model that are implemented as neural networks. Specifically, \textcolor{orange}{QR} indicates that $\eta_\text{self}$ is a neural network, \textcolor{orange}{Belief} signifies that $\eta_\text{other}^s$ is a neural network, and \textcolor{orange}{L} means that a neural network predicts a mixture of level-$k$ players. The numbers in parentheses represent the standard errors from the 10-fold cross-validation.
    \label{tab:robustness_check}
\end{table}

\section*{Experimental Instructions and Comprehension Checks}

\begin{figure}[h!]
\centering
\includegraphics[width=.9\textwidth]{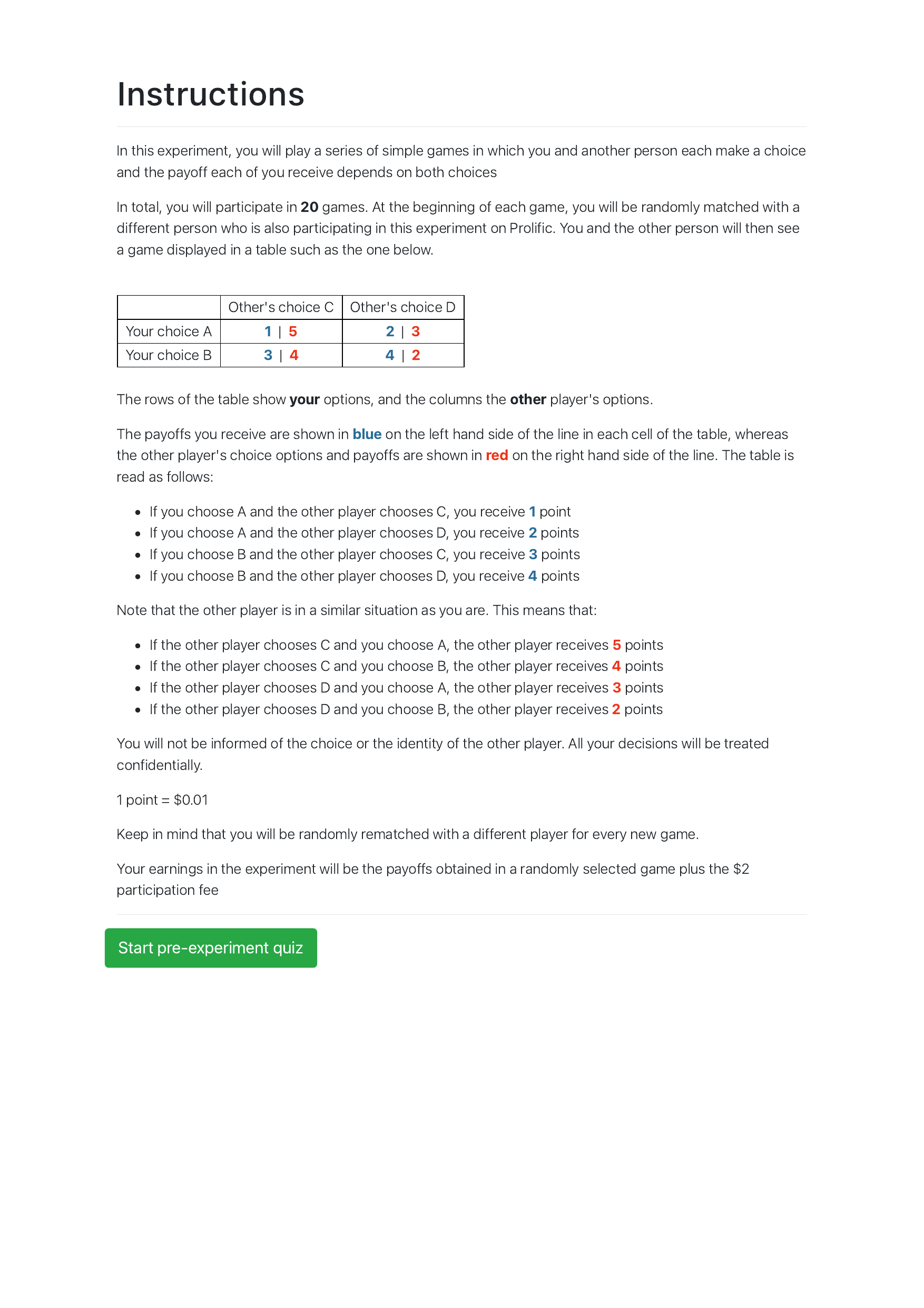}
\label{fig:S1_instructions}
\end{figure}
\noindent {\bf Fig. S1.} Screenshot of experimental instructions presented to participants in the main experiment.

\begin{figure}[h!]
\centering
\includegraphics[width=.9\textwidth]{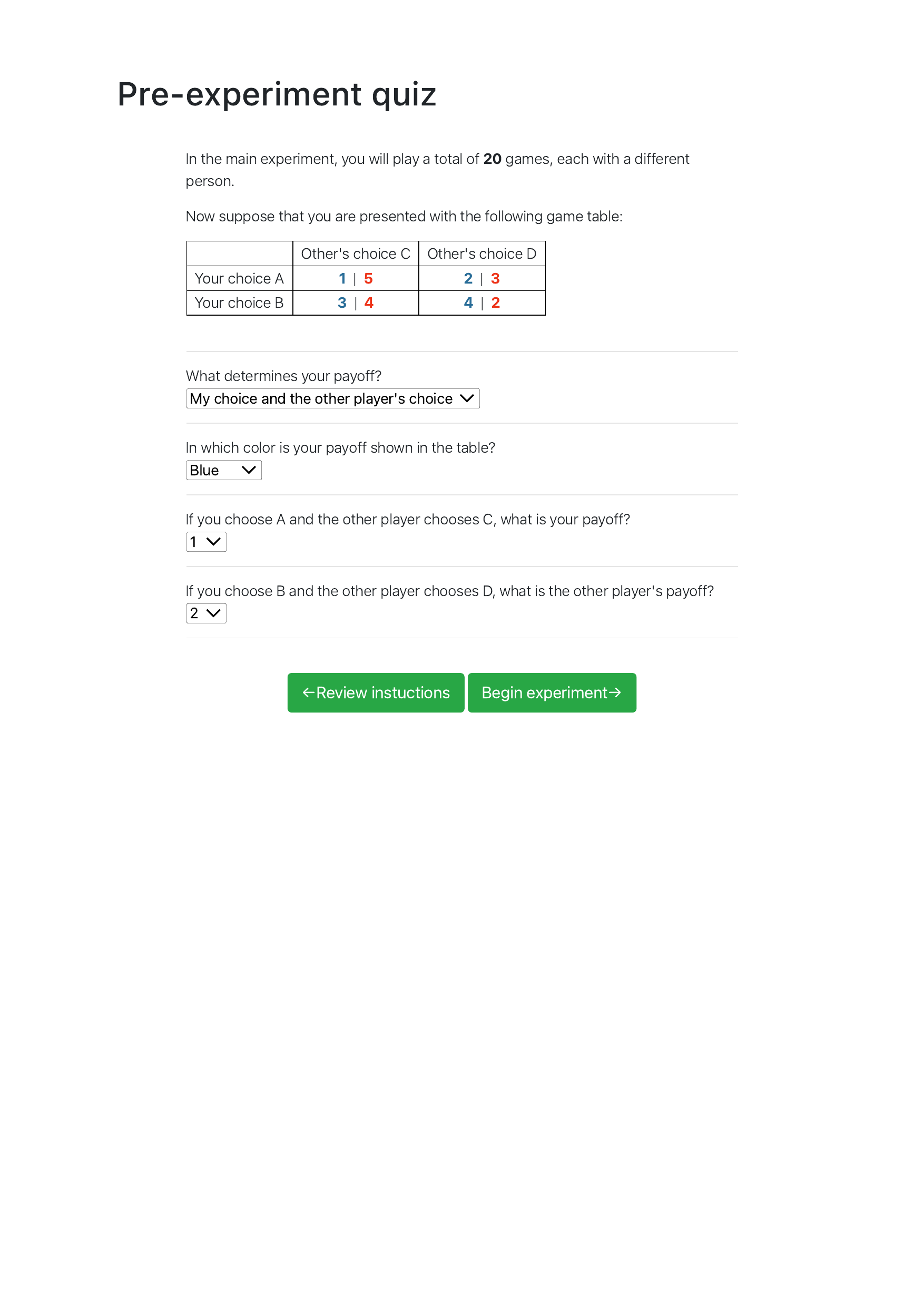}
\label{fig:S2_understand_check}
\end{figure}
\noindent {\bf Fig. S2.} Screenshot of the comprehensive check in the main experiment.

\begin{figure}[h!]
\centering
\includegraphics[width=.9\textwidth]{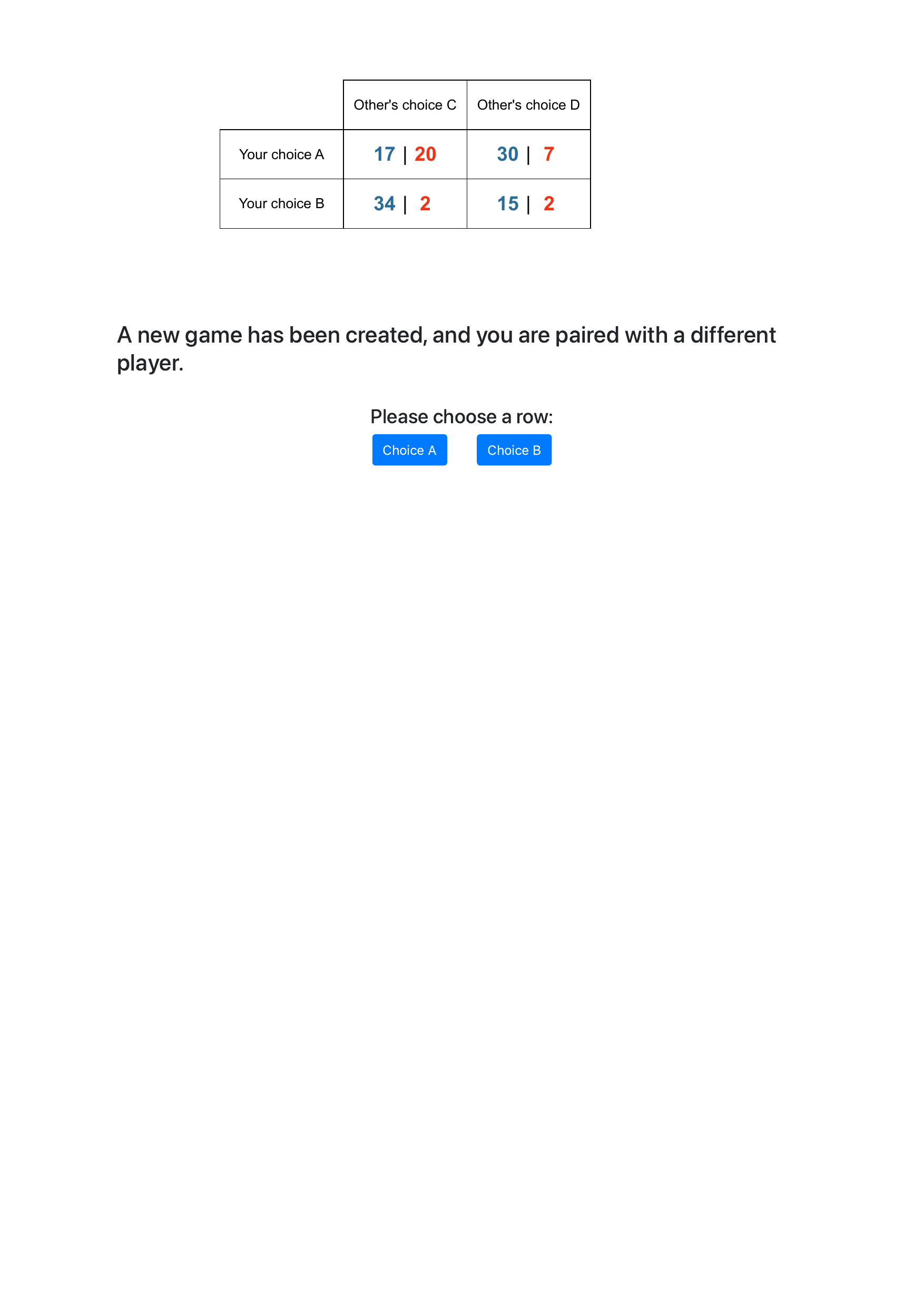}
\label{fig:S3_main_exp}
\end{figure}
\noindent {\bf Fig. S3.} Screenshot of a trial showing the game matrix to a participant in the main experiment.


\begin{figure}[h!]
\centering
\includegraphics[width=\textwidth]{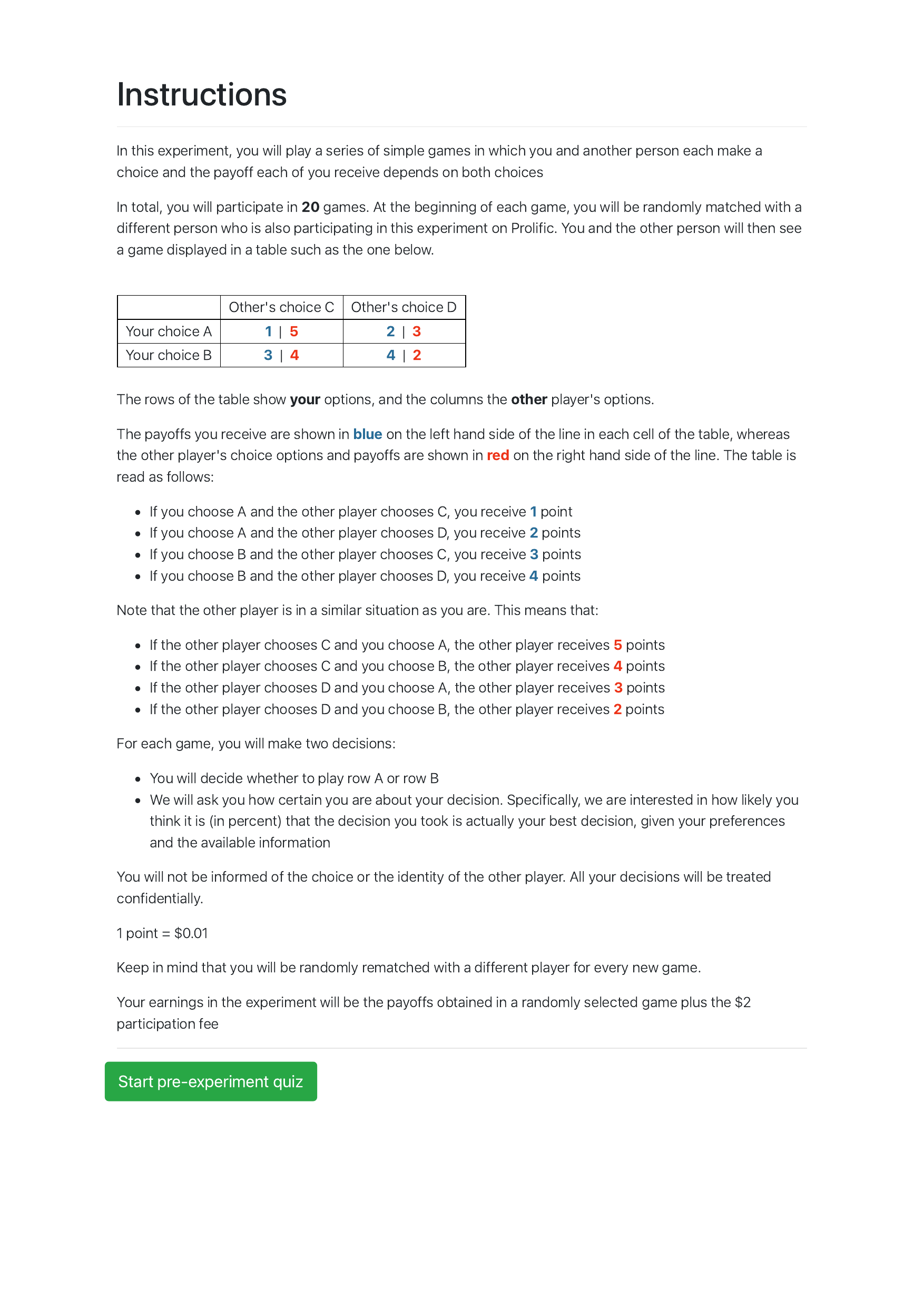}
\label{fig:S5_exp2_instructions}
\end{figure}
\noindent {\bf Fig. S4.} Screenshot of experimental instructions presented to participants in the follow-up experiment.


\begin{figure}[h!]
\centering
\includegraphics[width=\textwidth]{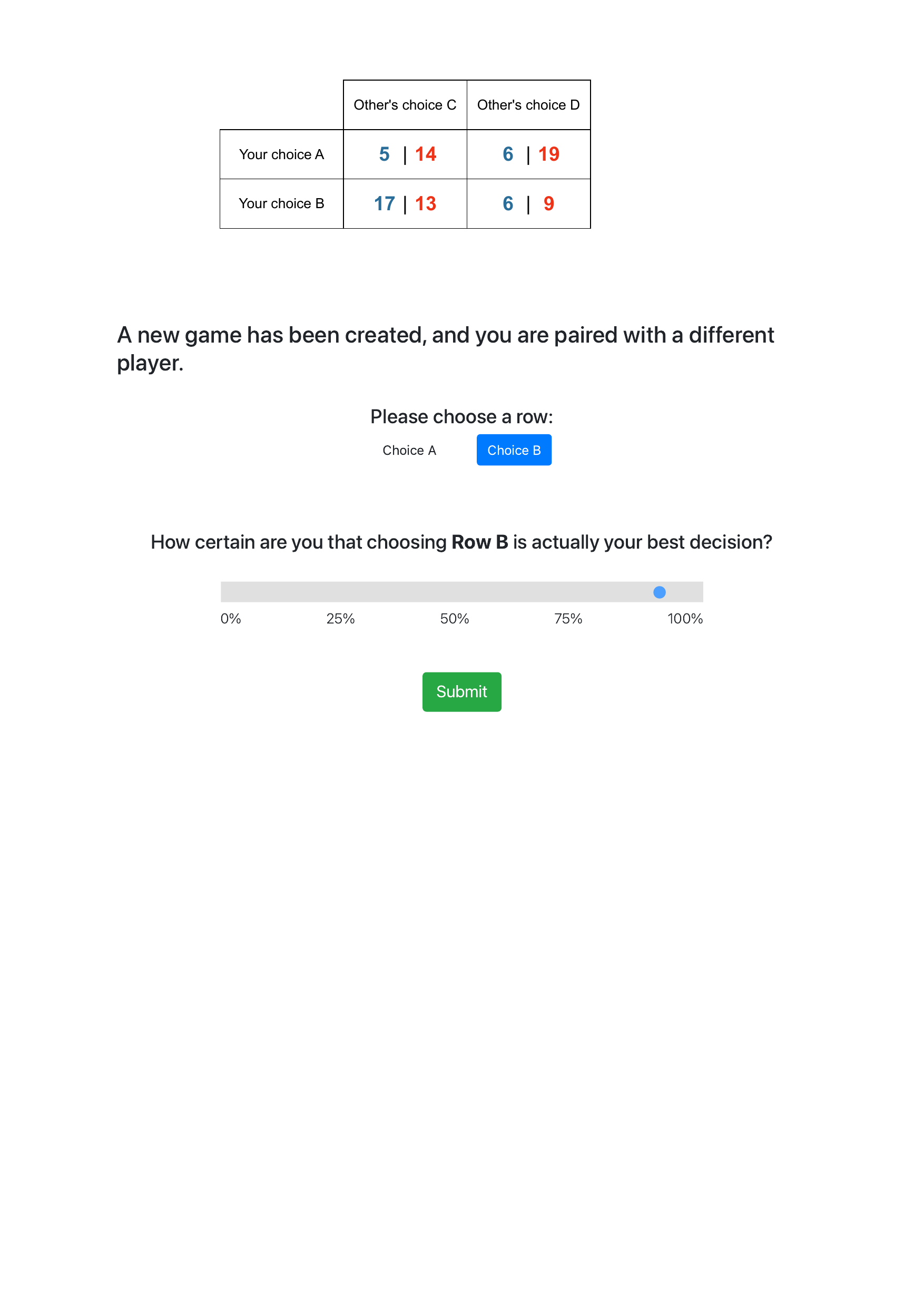}
\label{fig:S5_exp2_instructions}
\end{figure}
\noindent {\bf Fig. S5.} Screenshot of a trial showing the game matrix and confidence rating to a participant in the follow-up experiment.

\end{appendices}

\end{document}